\documentclass[journal]{IEEEtran}

\usepackage{graphicx,amssymb,lineno}
\usepackage{amsmath,amsfonts,amssymb}

\usepackage[usenames]{color}
\usepackage{float}
\usepackage{epstopdf}
\usepackage{longtable,booktabs}
\usepackage{mathtools}

\usepackage{multirow}
\usepackage{longtable}
\usepackage{graphicx,graphics,color,epsfig,subfigure,graphpap,rotate}
\usepackage{times,verbatim,subfigure,epsfig,graphicx,latexsym,amsmath}
\usepackage{url,cite}
\usepackage{subfigure}

\makeatletter
\newif\if@restonecol
\makeatother

\usepackage[linesnumbered,ruled,vlined]{algorithm2e}
\usepackage{algpseudocode}
\usepackage{amsmath}

\begin{document}
%
\title{Scheduling of UAV-assisted Millimeter Wave Communications for High-Speed Railway}

\author{Yibing~Wang,
        Yong~Niu,~\IEEEmembership{Member,~IEEE},
        Hao~Wu,
        Shiwen~Mao,~\IEEEmembership{Fellow,~IEEE},
        Bo~Ai,~\IEEEmembership{Fellow,~IEEE},
        Zhangdui~Zhong,~\IEEEmembership{Fellow,~IEEE},
        and Ning~Wang

\thanks{Y. Wang, H. Wu, B. Ai and Z. Zhong are with the State Key Laboratory of Rail Traffic
 Control and Safety, Beijing Jiaotong University, Beijing 100044, China (Email:
 18111034@bjtu.edu.cn).}
\thanks{Y. Niu is with the State Key Laboratory of Rail Traffic Control and Safety,
Beijing Jiaotong University, Beijing 100044, China, and also with the National
Mobile Communications Research Laboratory, Southeast University, Nanjing
211189, China (Email:niuy11@163.com).}
\thanks{S. mao is with the Department of Electrical and Computer Engineering,
Auburn University, Auburn, AL 36849-5201, USA (Email: smao@ieee.org).}
\thanks{N.~Wang is with the School of Information Engineering,
Zhengzhou University, Zhengzhou 450001, China (Email: ienwang@zzu.edu.cn).}
}

\maketitle

\begin{abstract}
To exploit richer spectrum resources
for even better service quality, millimeter wave (mmWave) communication
has been considered
for
high-speed railway (HSR) communication systems. In this paper, we
focus on scheduling
as many flows as possible
while satisfying
their QoS requirements.
Due to interference, eavesdropping, or other problems,
some flows may not be
directly transmitted from the track-side
BS.
In this paper, we propose an UAV-assisted scheduling scheme which utilizes
a
UAV to serve as relay for such flows.
The proposed scheme
also utilize two mmWave bands,
one
for the BS links
and the other for the UAV links.
The proposed algorithm aims to maximize
the number of flows with their QoS requirements satisfied.
Simulations demonstrate that the proposed scheme
achieves a
superior performance on the number of completed flows and the system throughput over two baseline schemes.
\end{abstract}


\section{Introduction}\label{S1}

With the rapid development of high-speed railway (HSR) in recent years,
there is a compelling need for high-quality broadband wireless access services for HSR.
There is increasing demand to support data- and bandwidth intensive applications (e.g., multimedia applications) for HSR passengers.
The current transmission rate requirement of each train cabin is about 37.5 Mbps according to a recent study, while
may continue to grow and reach $0.5\thicksim5$ Gbps in the future~\cite{HSR}. Obviously, these current transmission schemes cannot meet such increasing demands.

To this end, the millimeter wave (mmWave) band from 30 GHz to 300 GHz has abundant spectrum resources and can provide multi-gigabit transmission rates for
bandwidth-intensive
services,
such as
high-speed data transmission between devices (such as cameras, smartphones, tablets and laptops), compressed and uncompressed high-definition TV, real-time playback, wireless gaming, wireless personal area network, cellular access and wireless backhaul, etc.~\cite{mm,HD,HD1,mm1}.

In this paper, we consider the mmWave communications
between the HSR and the ground base station (BS).
To overcome
the severe penetration loss of the train body shell, multiple mobile relays (MRs) are deployed on the rooftop of the train
to provide better connections to the track-side BS.
The passengers in the train communicates with the BS through radio access links,
using the MRs as relays.
MRs serve passengers via the APs which are installed in each carriage to avoid the penetration loss and frequent handover.
Note that,
the
communications inside the train can
use
multiple radio access technologies,
such as LTE or WiFi.
In fact, the links between the MRs and the track-side BS are the main cause of the capacity bottleneck~\cite{HSR2}. Thus, we focus our work on the mmWave communications between the BS and MRs.

In the typical scenario of
mmWave HSR systems, there are a certain number of traffic
flows to be transmitted.
Each flow has its own minimum throughput requirement (i.e., the lowest transmission rate requirement), which is also referred to as its quality of service (QoS) requirement in this paper. Due to the diversity of applications, the
QoS requirements of different flows are
diverse. Furthermore,
the mutual interference from concurrent link transmissions and the presence of eavesdroppers
may cause some flows fail to be scheduled directly~\cite{MR}.
Therefore, we propose to
deploy a UAV to relay the flows that cannot be directly transmitted from the BS to MRs under certain
conditions.
We present an UAV-assisted scheduling algorithm
that exploits two mmWave bands (including one lower band for links from the BS and
a higher band for links from the UAV) to achieve
high
system throughput.
We also
consider
protecting the mmWave communications from potential eavesdroppers by
enforcing the security capacity
constraint.
The main contributions of this paper
are
summarized as follows.

\begin{itemize}
  \item
  To increase the network capacity of the HSR communication system and satisfy the needs of HSR passengers, we exploit the available spectrum resources of two mmWave frequency bands for the scheduled
  flows with diverse
  QoS requirements. The lower one of the two mmWave frequency bands is used
  for links from the BS, and the
  higher one is used
  for links from the
  UAV. Besides, we consider the existence of potential eavesdroppers, we
  determine
  whether a
  flow can be scheduled based on the security capacity to ensure its communication security.

  \item Due to interference, eavesdropping, or other problems, some flows cannot be directly scheduled from the BS to the
  MRs.
  To address this issue,
  we propose to use
  a
  UAV as a
  relay. With the
  assistance
  of the UAV,
  there is more flexibility for scheduling the flows.
  We formulate the problem of optimal scheduling as a nonlinear integer programming problem. Moreover, we propose a heuristic UAV-assisted scheduling algorithm to maximize the number of flows with their QoS requirements satisfied within a fixed time.

  \item
  We evaluate the proposed UAV-assisted scheduling scheme for the mmWave HSR network with
  simulations. Compared with two baseline schemes, the simulation results demonstrate that the proposed scheme
  achieves
  a
  superior performance
  on both the number of completed flows and the system throughput.
\end{itemize}

The remainder of this paper is organized as follows. In Section~\ref{S2}, we review the related work. Section~\ref{S3} presents the system model and assumptions.
In Section~\ref{S4}, we formulate the optimal scheduling problem as a nonlinear integer programming problem. In Section~\ref{S5}, we propose
the relay decision algorithm and UAV-assisted scheduling algorithm. Section~\ref{S6} presents
our simulation
study of
the proposed scheme.
We conclude this paper in Section~\ref{S7}.

\section{Related Work}\label{S2}

There are significant efforts on exploiting mmWave communications for
augmented capacity
for HSR systems.
For the problem of flow forced interruption,
utilizing other receivers/nodes as a one-hop or multihop relay to change
the original transmission path
is the most common solution~\cite{relay,relay1,relay2}.
In this paper, we do not use any original nodes to relay blocked flows,
but use the UAV to create a new,
high-altitude relay node. Due to the superiority of the UAV's open location,
superior relay effects can usually be achieved.
Many prior works
use alternative APs, which are densely deployed,
to substitute direct paths from transmitter to
receiver~\cite{AP,AP1,AP2,Gao}. In this
work, we adopt alternative APs in the HSR system,
i.e., the
MRs are deployed on
top of the train to
assist transmissions from the BS to
passengers in the train.

There have been considerable prior works on scheduling in mmWave networks.
Time division multiple access (TDMA) based schemes are commonly used in mmWave networks.
Such schemes
consider different types of link transmissions, including radio access, backhaul,
and so on~\cite{TDMA,TDMA1,QD,HeTVT17}. At present,
many existing schemes
aim to satisfy
the QoS requirements of flows, where
the scheduling scheme is designed under the premise of meeting the QoS requirement of each flow, and aiming to finally achieve different optimization goals~\cite{QoS,QoS1}. In this work,
the goal is to develop a
QoS-aware solution to schedule flows in HSR systems, which is assisted by an UAV relay.
We consider the cooperation of two mmWave bands to achieve greater transmission
capacity.
We also consider the corresponding security measures
against the communication security risks from
potential eavesdroppers.

Recently, there have been a few
studies on UAV-assisted relay networks~\cite{UAV-IoT21}.
The scheme of the optimal UAV placement is proposed for the single user or multiple users in the maritime communication, which can increase the capacity of wireless backhaul link between BS and UAV~\cite{ad1}. A cooperative secrecy transmission mechanism is presented for multiple UAV assisted relay selection to maximize the secrecy capacity~\cite{ad3}. In this paper, the transmission research is in HSR systems with the specific communication environment, and the completed link number and system throughput are maximized by scheduling link transmissions.
UAV can
serve
as an energy source and provide RF energy for low power device-to-device (D2D) pairs~\cite{UAV}. Moreover,
UAV enables
the simultaneous wireless information and power transfer (SWIPT) system by jointly optimizing time switching-based relaying (TSR) and
the optimization of its hovering location for better channel quality to maximize the network throughput~\cite{UAV1}.
These works usually assume the Rician fading channel
as well as the line-of-sight (LoS) path between the UAV and each receiver~\cite{UAV2}.
To better model
the actual channel state in
HSR systems
and to provide a better fit to real measurements observed in LoS scenarios, the
$\kappa$-$\mu$ distribution was proposed in~\cite{UAV3,UAV4,UAV5,UAV6} as a general fading channel model where the Rician, Nakagami-m, and Rayleigh distributions are its special cases.
In this paper, we adopt the
$\alpha$-$\beta$ path loss model which is
more in line with the path loss measurements
in the HSR scenario.
Besides, there are works which
have studied the application of UAV-assisted
communications
in HSR system. Two UAV-assisted relaying transmission schemes are presented to allocate limited power for moving source and UAV, which are to save signaling overhead during the entire transfer process~\cite{ad2}. In this work, we proposed QoS-
aware solution to maximize the number of scheduled flows. In addition, we allocate the time resource for two types of links and consider the cooperation of two
mmWave bands for the links from BS to UAV and the links from UAV to MRs.

It is worth noting that
these existing related works have not considered utilizing UAV to relay
traffic
flows which cannot be directly transmitted from the BS to MRs.
In this paper,
we integrate the
dual-band cooperation of mmWave communications into the scheduling problem, aiming to maximize the number of scheduled flows.
We also aim to satisfy the QoS requirements of each flow and avoid the potential risks caused
by eavesdroppers.
We will show through analysis and simulation that the proposed scheme can achieve a superior performance compared to the baseline schemes.

\section{System Overview}\label{S3}


\subsection{System Model}\label{S3-1}

\begin{figure*}[!t]
  \begin{center}
  \includegraphics[width=16cm]{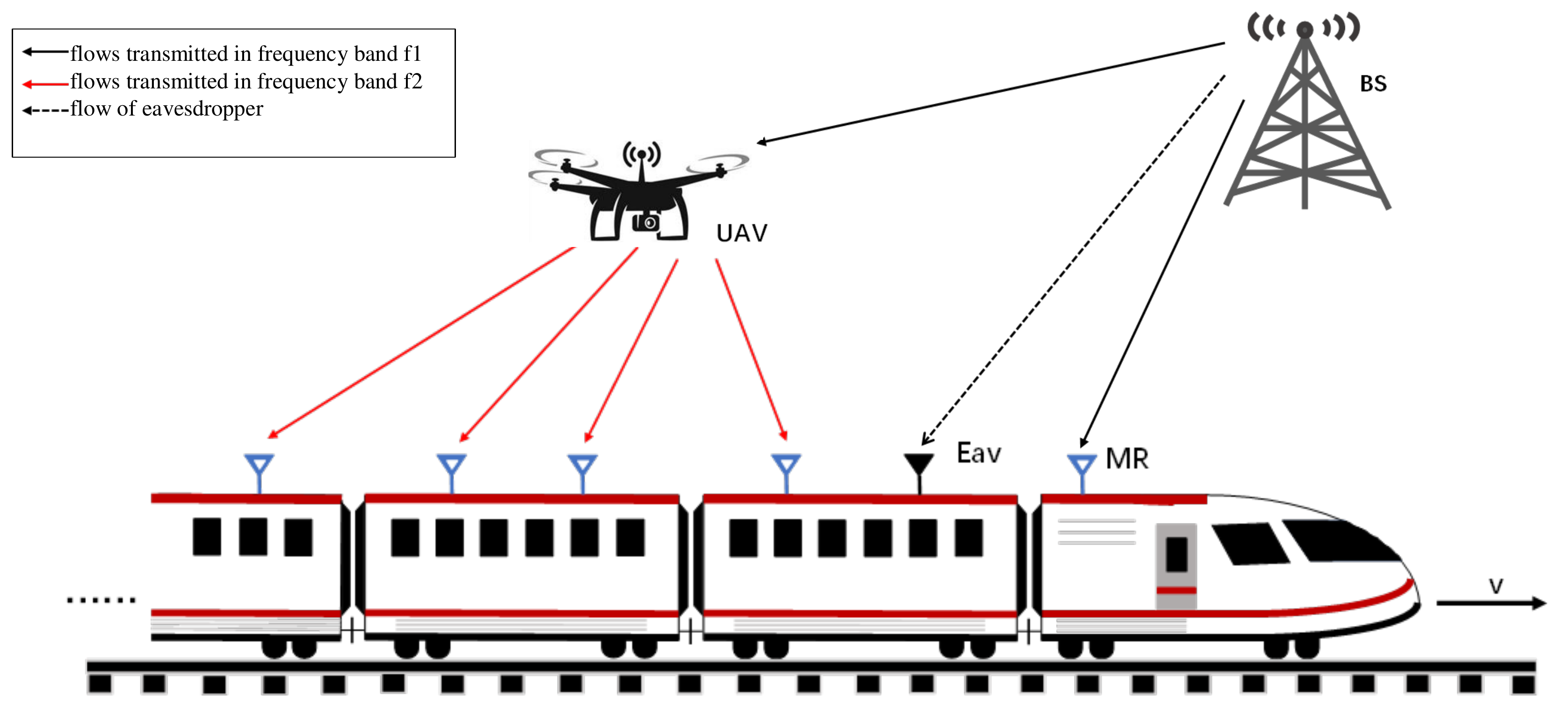}
  \end{center}
  \caption{The mmWave HSR communication system.}
  \label{fig:HRS}
\end{figure*}

In this paper, we consider a single cell scenario of the mmWave HSR communication
system, as
shown in Fig.~\ref{fig:HRS}. A fixed BS is located
beside the railway track, and
a UAV is hovering above at a fixed position
within the coverage of the track-side BS.

which are within
the coverage range of the BS and UAV.
The BS links are on frequency band $f_1$ , indicated by the
black arrows in Fig.~\ref{fig:HRS}. The communications between the UAV and MRs are
on
frequency band $f_2$, which are shown as
red
arrows in the figure.
In addition, there is an eavesdropper, termed $Eva$, which is trying to overhear
the communications. Without loss of generality, we assume
$Eva$ can only eavesdrop on the BS through frequency band $f_1$.

There are $F$ traffic flows between the BS and MRs, which need to be scheduled in the system.
The UAV provides relay service in the air, thus each flow may be received from
the BS directly or be relayed by the UAV
for the destination MR.
MRs can communicate with the track-side BS or the aerial UAV through radio access links. According to the selection of the transmission methods of the
flows,
some links are scheduled for transmission each time.
In this paper,
a
link
represents the actual connection of direct communications between two endpoints in the system.

In this paper, we consider the scenario where a train runs at a constant velocity $v=300$km/h. In this case, the speed of the
train is in line with reality. Although the train is moving at a high speed, the large bandwidth of mmWave allows the train to receive a large amount of data in a very short time. Therefore, the pressure of handover problem in high-speed railway communication can be greatly alleviated.
We
consider the downlink transmissions.
The auxiliary transmission of UAV can effectively alleviate the
link blockage
problem caused by obstacles
on the ground. All the
devices
in the system are assumed to operate in the
half-duplex
mode, i.e.,
they can only transmit or receive data at a time.
Each MR
is
equipped with one steerable directional antenna, so it
can only receive content from the BS or UAV at
a
time. The UAV is equipped with one directional
antenna for receiving and $N_{U_t}$ antennas for transmitting. The BS is equipped with $N_{B_t}$ directional antennas for transmitting.

In the proposed
system, time is divided into a series of non-overlapping super frames.
Each super frame consists of a scheduling phase and a transmission phase. The scheduling phase is the time duration to collect
link requests and their QoS requirements.
The transmission phase consists of $M$ equal time slots (TS). For clarity of exposition,
we use TSs to measure the transmission time. For reliable transmissions, the control messages
and BS requests can be
transmitted with 4G technologies, including
the source/destination
information and the QoS requirements of each link~\cite{4G}.
The proposed scheduler is to allow multiple links be transmitted concurrently
using
spatial-time division multiple access (STDMA)~\cite{Qiao}.

\subsection{Antenna Model}\label{S3-4}

As in prior works, directional antenna is necessary
for mmWave communications~\cite{antf}.
We
assume
that the
BS and UAV are both equipped with multiple directional antennas,
which means they can send to multiple links simultaneously.
However, each MR
can receive
from only the BS or UAV at a time.
We adopt multiple antennas to form multiple directional beams at the BS and UAV for transmitting multiple data flows
simultaneously~\cite{ant0}. The maximum numbers of beams formed simultaneously at
the
BS and UAV are limited by the numbers of theirs equipped transmitting antennas.

In this paper, we adopt the realistic directional antenna model which is mentioned in \cite{23}.
All directional antennas can steer their beams
for the maximum directivity gain~\cite{ant1, ant2}. The perfect beam alignment for the desired link has the maximum directivity antenna gain of $G_0$. The antenna model can be expressed as
\begin{equation}\label{antenna}
G_a(\theta)=
\begin{cases}
G_0-3.01\cdot (\frac{2\theta}{\theta_{-3dB}})^2, &{0^{\circ}\leq \theta \leq \theta_{ml}/2} \\
G_{sl}, &{ \theta_{ml}/2\leq \theta \leq 180^{\circ}}
\end{cases}
\end{equation}
where $\theta$ is an angle within the range $[0^{\circ}, 180^{\circ}]$. $G_0$ is the maximum antenna gain, $G_0 = 10\log(1.6162\sin(\theta_{-3dB}/2))^2$. $\theta_{-3dB}$ is the angle of the half-power beamwidth. $\theta_{ml}$ is the main lobe width, $\theta_{ml}=2.6\cdot\theta_{-3dB}$. $G_{sl}$ is the sidelobe gain, $G_{sl} = -0.4111log(\theta_{-3dB})-10.579$.

The directional antenna model is for the antenna gain calculations of the links.
%

\subsection{Transmission Model}\label{S3-2}

The directional path loss in the target environment
is studied via the impacts of antenna beam and propagation conditions.
After evaluating
the different existing
models, the
$\alpha$-$\beta$
model is selected as the path loss model for quantitative analysis in this work~\cite{a-b}. For a link $i$, the path loss between its transmitter $t_i$ and its receiver $r_i$ is
\begin{equation}
PL(d_{t_ir_i}) = \alpha+10\beta \log10(d_{t_ir_i})+X_\sigma, \label{equation:PL}
\end{equation}
where
$\alpha$ and $\beta$ are constants, $d_{t_ir_i}$ is the distance between $t_i$ and $r_i$,
and $X_\sigma$ is a zero mean Gaussian random variable with a standard deviation $\sigma$.

Compared with line-of-sight (LOS) transmissions, non-line-of-sight (NLOS) transmissions suffer from much higher attenuation in the
mmWave band~\cite{attenuation}.
Therefore we only consider the directional LOS links.
For a link $i$, the received signal power at
receiver $r_i$ from
transmitter $t_i$ is
\begin{equation}
{P_r}(i,i) = G_{t_i} + G_{r_i} + Pt_i - PL(d_{t_ir_i}), \label{equation:Pr}
\end{equation}
where $G_{t_i}$ is the transmitter antenna gain of link $i$, $G_{r_i}$ is the receiver antenna gain of link $i$, $Pt_i$ is the
transmit
power at the
transmitter of link $i$, and $PL(d_{t_ir_i})$ is calculated
by~\eqref{equation:PL}.

Since we allow multiple links
transmit
concurrently by the STDMA transmission scheme, there
is mutual interference
between concurrent links
using the same frequency band.
For co-band concurrent links $i$ and $j$,
the interference at $r_i$ from $t_j$ can be calculated as
\begin{equation}
{I_{ji}} = G_{t_j}+ G_{r_i} +{Pt_j}-PL(d_{t_jr_i}), \label{equation:I}
\end{equation}
where $G_{t_j}$ is the antenna gain at the transmitter of link $j$, $G_{r_i}$ is the antenna gain at the receiver of link $j$, which is calculated as in~\eqref{antenna}, $Pt_j$ is the
transmit
power at the transmitter of link $j$, and $PL(d_{t_jr_i})$ is calculated by~\eqref{equation:PL}.

According to
Shannon's formula,
the transmission rate of link $i$ is given by
\begin{equation}
R_i=\eta W_i  \log_2 \left( 1+\frac{{P_r}(i,i)}{N_0W_i+\sum_{j \neq i} I_{ji}} \right),\label{equation:rate}
\end{equation}
where $\eta$ describes the efficiency of the transceiver design,
which in the range of $(0,1)$; $W_i$ is the transmission bandwidth of
link $i$; and $N_0$ is the one-sided
power spectra density of the white Gaussian noise.

We denote ${R_i^t}$ as the actual transmit rate of link $i$ in time
slot $t$, which is calculated by~\eqref{equation:rate}. Then the achieved throughput of link $i$
in this time slot
can be obtained as
\begin{equation}
{q_i^a} = \frac{{\sum_{t=1}^M{R_i^t} \cdot \Delta T }}{{T_s+ M\Delta T }}, \label{equation:qi}
\end{equation}
where
$T_s$ is the duration of the scheduling phase, and $\Delta T$ is the
duration of each TS.
The denominator in~\eqref{equation:qi} is the duration of a super frame.

\subsection{Secrecy Capacity Model}\label{S3-3}

In this paper,
we consider the existence of
an eavesdropper, which can
overhear
the BS transmissions
on frequency band $f_1$.
It is important to ensure
secure communications and prevent information leakage. The secrecy capacity, an important physical layer security performance
metric, is adopted
to determine the feasibility of the schedule scheme in this work.
Let $C_M$ and $C_E$
denote
the Shannon capacities of the main channel
($C_M = R_i$ if link $i$ is attacked)
and
the eavesdropping channel, respectively.
The secrecy capacity is given by~\cite{s-c}
\begin{equation}\label{secrecy capacity}
{C_S} = C_M-C_E,
\end{equation}
where $C_M$ and $C_E$ are both calculated by~\eqref{equation:rate}.

The position of the eavesdropper is constantly changing with the train moving. But the eavesdropper is fixed in a certain position of the train and is stationary relative to the train. Hence, compared with other randomly moving eavesdroppers, the location of the eavesdropper in this case is easier to estimate. The eavesdropper is not completely passive, and sends eavesdropping messages to the unknown receiver. The train is slow when it starts to run and can be regarded as stationary. At this time, we adopt the angle of arrival (AOA) positioning method to determine the specific location of the eavesdropper on the train~\cite{AOA}. The BS and MRs detect the signals sent by the eavesdropper to obtain the angles of incidences. Then the location of the eavesdropper can be determined by these angles of incidences. However, determination of the eavesdropper location facilitates the calculation of the security capacity.


\section{Problem Formulation And Analysis}\label{S4}

\subsection{Problem Formulation} \label{S4-1}

In this paper, the goal is to
accommodate
as many flows as possible through effective scheduling.
In fact, different flows carry different types of service.
Thus each flow $f$ has its
QoS requirement $q_f$
in the form of its required
minimum throughput.
We define a binary variable $\delta_f$ to indicate whether flow $f$
is scheduled successfully. If that is the case, we have $\delta_f=1$; otherwise, $\delta_f=0$.
Let there be $F$ flows. Then the objective function of
the scheduler is given by
\begin{equation}
\text{max} \sum_{i=1}^F \delta_f.\label{equation:object}
\end{equation}

For link $i$, a binary variable ${a_i^t}$ is defined to indicate whether
the link is scheduled for transmission in time slot $t$. If so, ${a_i^t=1}$; otherwise, ${a_i^t=0}$.
For the links generated from relayed flows, they are all transmitted to or received from the UAV.
The following constraint enforces the half-duplex transmission condition.
\begin{equation}
a_i^t+a_j^t\leq 1,  \;\;  \text{if}\ t_i = r_j \; \text{or} \; \ r_i=t_j. \label{equation:con1}
\end{equation}
The constraint means the link to the UAV and the link from the UAV can't be scheduled concurrently.
Each MR and the UAV can only receive a single link at a time.
In other words, the receivers of any two concurrent links $i$ and $j$ must be different. Thus, we have the following constraint.
\begin{equation}
r_i\neq r_j,  \;\; \text{if}\ a_i^t=1 \; \mbox{and} \; \ a_j^t=1. \label{equation:con2}
\end{equation}

Each link $i$ has its QoS requirement $q_i$.
The actual achieved throughput of each flow must satisfy its
QoS requirement.
For link $i$, the
QoS constraint is as follows.
\begin{equation}
q_i^a = \frac{{\sum_{t=1}^M{R_i^t} \cdot \Delta T }}{{T_s+ M\Delta T }}\geq q_i. \label{equation:con3}
\end{equation}

Regardless of whether it
be the BS or UAV, the numbers of links
can be
simultaneously
transmitted
are
limited by the numbers of its
transmit
antennas. In time slot $t$,
the set of links transmitted from
the
BS is defined as $S_B^t$, and the set of links transmitted from
the
UAV is defined as $S_U^t$.
Then we have the following hardware constraint.
\begin{equation}
\sum_{i\in S_B^t}a_i^t\leq |N_{B_t}|, \;\;
\mbox{and} \;\;  \sum_{j\in S_U^t}a_j^t\leq |N_{U_t}|. \label{equation:con4}
\end{equation}

Some flows are relayed by
the
UAV
and then forwarded to the target MRs.
The total throughput of the links transmitted from
the
BS is larger than the sum throughput of the links transmitted from
the
UAV.
\begin{equation}
\sum_{i\in \sum_{t=1}^\tau S_B^t}q_i^a\geq \sum_{j\in \sum_{t=1}^\tau S_U^t}q_j^a, \;\; \tau \leq M, \label{equation:con5}
\end{equation}
where $q_i^a$ and $q_j^a$ are calculated as in~\eqref{equation:qi}.

Due to the existence of an eavesdropper, the confidentiality of transmissions must be guaranteed. The eavesdropping channel between a legitimate pair
is
a degraded version of the main channel~\cite{s-c}. In time slot $t$, the main
channel capacity and the eavesdropping
channel capacity of link $i$ are denoted by $C_M^i$ and $C_E^i$,
with the following constraint~\cite{s-c1}.
\begin{equation}
C_E^i<0.1\cdot C_M^i, \;\; i\in S_B^t, \label{equation:con6}
\end{equation}

The problem of maximizing the amount of scheduled flows in
in each time slot can be formulated as follows.
%
\begin{align}
({\rm{P1}}) \ \max & \; \sum_{i=1}^F \delta_f, \label{OBJ} \\
\mbox{s.t.} & \; \eqref{equation:con1}-\eqref{equation:con6}. \nonumber
\end{align}
It can be seen that
Problem P1 is a nonlinear integer programming problem.
There are also complex nonlinear terms in the constraints. The problem is actually
NP-hard~\cite{QoS,sch}.
Thus, we aim to develop heuristic algorithms
to obtain competitive solutions with low computational complexity.

\section{The UAV-assisted Scheduling Scheme}\label{S5}

In this section, we propose an UAV-assisted
scheduling scheme. With this scheme, each flow
can choose from two transmission methods: one is to directly transmit from
the
BS to
the target
MR, and the other is to
use the UAV as a relay.
The proposed scheme
first determines the transmission method for each flow
and then schedules
the links to transmit, aiming to
maximize the number of scheduled flows.

\subsection{The Relay Decision Algorithm}\label{S5-1}

The relay decisions of the flows are based on their QoS requirements and the transmission efficiency.
The flows that need to be scheduled are determined based on the transmitter-receiver ranges and their QoS requirements.
Then we schedule the flows in each time slot for transmission.

We first consider whether the QoS requirements of the flows
can be
satisfied. Then we make a selection for each flow between being directly transmitted and being relayed.
The set of the links which select to be directly transmitted from BS to MR is defined as $S_1$, and the set of the links relayed by the UAV is defined as $S_2$.

For each flow $f$, if it is directly transmitted from
the
BS to MR in the $f_1$ band, then this decision
will allow
one link $l_1$ from BS to MR to be scheduled. The achieved throughput $q_{l_1}^a$ of this link
is computed as in~\eqref{equation:qi}. If flow $f$
is to be
relayed by
the
UAV in the $f_2$ band,
then
there are two links $l_2$ and $l_2^\prime$ to be scheduled:
one from
the
BS to UAV and the other
from the
UAV to MR. The achieved throughput of these two links, $q_{l_2}^a$ and $q_{l_2^\prime}^a$,
can be calculated by~\eqref{equation:qi}. Since it is uncertain which links are transmitted simultaneously,
the above throughput calculated by~\eqref{equation:qi}
does not consider
the interference from
other links.
We can judge whether the two decisions (relay and no
relay) can successfully transmit the flows by comparing the calculated throughputs and the QoS requirements.
For flow $f$, the choice
between direct transmission and
relayed
transmission can be decided as follows.
\begin{enumerate}
\item If $\min \{q_{l_1}^a,q_{l_2}^a, q_{l_2^\prime}^a\} \geq q_f$, flow $f$ can be scheduled by both
choices.
\item If $q_{l_1}^a \geq q_f$ and $\min \{q_{l_2}^a,q_{l_2^\prime}^a\}<q_f$, flow $f$ can only be directly transmitted from the
BS.
\item If $q_{l_1}^a<q_f$ and $\min \{q_{l_2}^a,q_{l_2^\prime}^a\}\geq q_f$, flow $f$ need to be relayed by the UAV.
\item If $q_{l_1}^a<q_f$ and $\min \{q_{l_2}^a,q_{l_2^\prime}^a\}<q_f$, flow$f$ should be
abandoned. It
should not be considered in the later scheduling process.
\end{enumerate}

Since the eavesdropper
is stealthily listening on the BS transmissions in the $f_1$ frequency band, the confidentiality of transmission should be considered. The location of
the
UAV
ensures that
the BS-UAV links
can satisfy constraint~\eqref{equation:con6}.
Thus the flows relayed by
the
UAV
are less likely of
being eavesdropped. For flows which are directly transmitted from
the
BS in
the
$f_1$ frequency band,
we judge whether a direct transmission is feasible
by evaluating~\eqref{equation:con6}.

To further compare the
two transmission choices, we make a preliminary estimation
of the transmission time of all the flows in
$S_{all}=S_1 \cup S_2$.
The number of estimated time slots of flow $f$ can be obtained as follows.
\begin{equation}\label{esttime}
Te_f=
\left\{
\begin{aligned}
&\frac{q_f}{R_{l_1}\Delta T} & & \text{if } f\in S_1\\
&\frac{q_f}{R_{l_2}\Delta T}+\frac{q_f}{R_{l_2^\prime}\Delta T} & & \text{if }  f\in S_2.
\end{aligned}
\right.
\end{equation}
Here $R_i$ is the
estimate rate without considering other links' interference at this time.
The
remaining flows select a
better transmission choice by comparing the number of TSs required by
the
schedule.

The relay decision algorithm is summarized in Algorithm~\ref{alg:selection_algorithm}.
First, we
remove
the flows
whose QoS requirements cannot be satisfied
by either of the two transmission choices in Line 2. In addition to the flows in set $\mathbb{D_f}$, the other flows are divided into three categories based on whether their
QoS requirements can be satisfied
under different transmission choices in Lines 5-24.
In Lines 5-6,
the
flows with their
QoS requirements satisfied by being relayed by the UAV
are added to $S_2$.
In
Lines 9-13,
the transmission methods for the
other flows are determined,
with their QoS requirements satisfied
by direct transmission from the
BS
and satisfying
the
secrecy capacity condition. And these flows will be put in $S_1$, the set of flows to be directly transmitted from the BS.
In Lines 14-24,
flows
whose QoS requirements can be satisfied by both transmission methods
will
be divided into three categories according to
the
secrecy capacity condition and transmission efficiency.
If the secrecy capacity condition cannot be satisfied,
it will be put in the set of flows relayed by the UAV, as in Lines 15-16.
Finally,
the
remaining flows select
their
transmission methods by comparing the required transmission time of different choices as in
Lines 19-24.

\begin{algorithm}[!t] 
	\small
	\caption {The Transmission Method Selection
	Algorithm} \label{alg:selection_algorithm}
	\KwIn{$S_{all}$, $S_1=\emptyset$, $S_2=\emptyset$, QoS requirements of flows, locations of the BS, UAV, eavesdropper and MRs }
	\KwOut{$S_1$, $S_2$}
	Calculate $q_{l_1}^a$, $q_{l_2}^a$ and $q_{l_2^\prime}^a$ for each flow\;
	Remove $\mathbb{D_f}=\{f|q_{l_1}^a<q_f \; \mbox{and} \min\{q_{l_2}^a,q_{l_2^\prime}^a\}<q_f\}$ from $S_{all}$\;
	\If{$|S_{all}|\neq 0$}{
		\For{flow $f$ ($1\leq f\leq |S_{all}|$)}{
			\If{$q_{l_1}^a<q_f$ and $\min\{q_{l_2}^a,q_{l_2^\prime}^a\}\geq q_f$}{
				$S_2=S_2\cup f$\;$S_{all}=S_{all}-f$\;}
			\uElseIf{$q_{l_1}^a\geq q_f$ and $\min\{q_{l_2}^a,q_{l_2^\prime}^a\}<q_f$}{
				\If{$C_E^{l_1}$ can satisfy~(\ref{equation:con6})}{
					$S_1=S_1\cup f$\;$S_{all}=S_{all}-f$\;}
				\Else {
					$S_{all}=S_{all}-f$\;}
			}
			\uElseIf{$\min\{q_{l_1}^a,q_{l_2}^a,q_{l_2^\prime}^a\}\geq q_f$}{
				\If{$C_E^{l_1}$ can't satisfy~(\ref{equation:con6})}{
					$S_2=S_2\cup f$\;$S_{all}=S_{all}-f$\;}
				\Else{
					\If{$\frac{q_f}{R_{l_1}\Delta T}\leq (\frac{q_f}{R_{l_2}\Delta T}+\frac{q_f}{R_{l_2^\prime}\Delta T})$}{
						$S_1=S_1\cup f$\;$S_{all}=S_{all}-f$\;}
					\Else{
						$S_2=S_2\cup f$\;$S_{all}=S_{all}-f$\;}
				}
			}
		}
	}	
\end{algorithm}

For the complexity of the
transmission method selection algorithm,
the
number of the iterations of the outer {\tt for} loop in Line 3 is $|S_{all}|$, where $|S_{all}|$ in the worst case is $\mathcal{O}(F)$. Hence, the computational complexity of the algorithm is $\mathcal{O}(F)$.

\subsection{The UAV-assisted Scheduling Algorithm}\label{S5-2}

Next we present
a transmission scheduling algorithm. This scheduling algorithm is developed
on the basis of the above transmission method selection algorithm.

We select the appropriate links for transmission
in each TS based on the principle of maximizing the number of transmitted flows.
For ease of exposition, we first introduce the simultaneous transmission requirements of links and the priority of link scheduling.
Regarding the contention among links, we consider the three cases as follows.
\begin{enumerate}
\item Due to half-duplex transmissions,
the link from the BS to the UAV and the link from the UAV to an MR cannot be scheduled
concurrently.
\item If the simultaneous transmissions of two links will cause at least one of them to fail to satisfy
the QoS requirement, there is a conflict between the two links.
\item As stated in~\eqref{equation:con6}, the links transmitted from the BS in the $f_1$ frequency band
that do not meet
the safe capacity threshold
will not
be scheduled simultaneously with other links.
\end{enumerate}

A
flow $f$ relayed by
the
UAV will require
two links $l_2$ and $l_2^\prime$ to be scheduled, i.e., $l_2$
from
the
BS to UAV and $l_2^\prime$
from the
UAV to MR. We define the set of links scheduled in
the
$f_1$ band as $S_{f_1}$, and the set of links scheduled in
the
$f_2$ band as $S_{f_2}$.
Link $l_2$
should
be scheduled before $l_2^\prime$.
Thus we first schedule link $l_2\in S_{f_1}$, and then link $l_2^\prime$ is generated in $S_{f_2}$ to wait for being scheduled.

Then we consider the transmit order of the links in $S_{f_1}$ or $S_{f_2}$. A parameter is defined to prioritize the scheduling order of the links, which is the inverse of a link's number of required TSs in a frame to satisfy its QoS requirement.
For
link $i$ which is generated from flow $f$, the priority value is given by
\begin{equation}\label{priority}
\varpi_i=\frac{R_i \Delta t }{q_f (T_s+M \Delta t)}.
\end{equation}
In fact, the less time
the transmissions consume, the more flows
can be
completed in a fixed time interval.
Prioritizing
the
flows that need
fewer
TSs can transmit more flows in a frame.
Therefore, we calculate the priority values of links in $S_{f_1}$ and $S_{f_2}$ as~\eqref{priority},
and then transmit
the links in the descending order in the scheduling process. Since there are always new links joining $S_{f_2}$, the scheduling order of the links in the set needs to be frequently
updated.

We define the set of the links which are to be
transmitted from
the
BS to UAV as $S_{BU}$, and a parameter $N_{f}$ is
used
to indicate the number of completed flows.
The QoS-aware
scheduling algorithm is summarized in Algorithm~\ref{alg:scheduling_algorithm}.
First, the input and output of the algorithm are specified.
Then, we configure sets $S_{f_1}$ and $S_{BU}$ in Lines 1-2.
In Lines 7-14,
we find the links in $S_{f_1}$ that can be scheduled simultaneously with other selected transmitted links.
In Lines 15-22,
the simultaneous scheduled links in $S_{f_2}$ are determined.
In Lines 23-40, we obtain the remaining throughput demands of all
the
scheduled links after this time slot, and examine whether there are flows which have finished transmission
in this TS.

\begin{algorithm}[!t]
	\small
	\caption {The UAV-assisted Scheduling Algorithm} \label{alg:scheduling_algorithm}
	\KwIn{$S_1$, $S_2$, $S_{BU}=\emptyset$, $S_{f_1}=S_1$, $S_{f_2}=\emptyset$, $N_f=0$, $SS=\emptyset$, $\alpha=0$, $n_{B_t}=0$, $n_{U_t}=0$, $q_f$}
	\KwOut{$N_f$}
	\For{flow $f\in S_2$}
	{
		$S_{f_1}=S_{f_1}\cup l_2$ (link $l_2$ is generated by flow $f\in S_2$ scheduling)\;
		$S_{BU}=S_{BU}\cup l_2$\;
	}
	Calculate the priority values of the links in $S_{f_1}$\;
	Sort the links in $S_{f_1}$ in increasing order by priority value\;
	\For{time slot $t$ $(1\leq t\leq M)$}
	{
		\If{$|S_{f_1}\cup S_{f_2}|\neq 0$}{
			\If{$|S_{f_1}|\neq 0$}{
				\For{link $i$ $(1\leq i\leq |S_{f_1}|)$}{
					\If{$n_{B_t}<N_{B_t}$}{
						\If{$i$ has no contention with the link(s) in $SS$}{
							$SS=SS\cup i$\;
							$n_{B_t}=n_{B_t}+1$\;
							\If{$i\in S_{BU}$}{
								$\alpha=1$\; $S_{f_2}=S_{f_2}\cup l_2^\prime$(link $l_2^\prime$ is generated by flow $f\in S_2$ scheduling)\;
							}
						}
					}
				}
			}
			\If{$\alpha=0$\&$|S_{f_2}|\neq 0$}{
				calculate the priority values of the links in $S_{f_2}$\;
				Sort the links in $S_{f_2}$ in increasing order by priority value\;
				\For{link $i^\prime$ $(1\leq i^\prime\leq |S_{f_2}|)$}{
					\If{$n_{U_t}<N_{U_t}$}{
						\If{$i^\prime$ has no contention with the link(s) in $SS$}{
							$SS=SS\cup i^\prime$\;
							$n_{U_t}=n_{U_t}+1$\;
						}
					}
				}
			}
		}
		\For{each link $j\in SS$}{
			Calculate the rate $R_j$ in the current time slot\;
			Calculate the remaining throughput demand of $j$, \\$q_j=q_j\ast(T_s+M\cdot\Delta t)-R_j^\delta\ast\Delta t$ ($q_j$ can obtain from the QoS requirement of the corresponding flow)\;
			\If{$q_j\leq 0$}{
				$SS=SS-j$\;
				\If{$j\in S_{f_1}$}{
					$S_{f_1}=S_{f_1}-j$\; $n_{B_t}=n_{B_t}-1$\;
					\If{$j\in S_{BU}$}{
						$\alpha=0$\; $S_{BU}=S_{BU}-j$\;
					}
					\Else{$N_f=N_f+1$\;}}
				\Else{
					$S_{f_2}=S_{f_2}-j$\; $n_{U_t}=n_{U_t}-1$\; $N_f=N_f+1$\;}
			}
		}
	}
\end{algorithm}

The outer {\tt for} loop
of the scheduling algorithm
has $\mathcal{O}(M)$ iterations. The {\tt for} loop in
Line 8 has $|S_{f_1}|$ iterations, and $|S_{f_1}|$ in the worst case is $\mathcal{O}(F)$. The {\tt for} loop in
Line 18 has $|S_{f_2}|$ iterations with the worst case value
$\mathcal{O}(F)$.
Moreover,
the {\tt for} loop in
Line 23 has $|SS|$ iterations, and its
worst case value
is also $\mathcal{O}(F)$. The {\tt for} loops in Line 8, Line 18 and Line 23 are parallel. Therefore, the complexity of the QoS-based scheduling algorithm is $\mathcal{O}(MF)$.
Such low complexity of the algorithm
makes it suitable
for practical implementation.

\section{Performance Evaluation}\label{S6}

In this section, we evaluate the performance
of the UAV-assisted scheduling algorithm.
The flow transmissions of the scheme involve two bands: $f_1$
in the 28GHz band and $f_2$
in the 60GHz band.
We also compare the performance of the proposed scheme with
several existing baseline
schemes.

\subsection{Simulation Setup}\label{S6-1}

We consider the
single cell scenario of the mmWave HSR system.
There are 24 MRs which are evenly distributed on
top
of the train. The flows which need to be scheduled are all transmitted from the BS. Then the flows are generated by randomly selecting the receivers among all the MRs.
The train has a total of eight cars with a total length of 200m.
Three MRs are
deployed
in each car.
One eavesdropper is randomly placed
on
top
of the train. The BS and UAV can reach a communication range of hundreds of meters, and all the
links in the system are transmitted within the allowable communication ranges of their transceivers. The number of flows which need to be transmitted between
the
BS and MRs is smaller than the number of MRs.

Since the MRs move at the high speed with the train, we need to update theirs location information periodically. We collect and update the location information of MRs every 2000 time slots in the process. During this period of time, the change distances of MRs do not exceed 3m, which does not affect the data transmissions.

Besides, each carriage of the train has 50-100 seats, then each MR has to deal with the data traffic requirements of 10-20 passengers at most. The flow of each MR request may be a combination of the traffic flows of multiple passengers. Thus, the QoS requirement values of scheduled flows fluctuates in a wide range. We assume that the QoS requirement of each flow is uniformly distributed between 10Mbps and 500Mbps. Other parameters are
presented in Table~\ref{tab:parameters}~\cite{a-b,ant0,Gao}.
For the path loss model,
the specific expression is
given in~\eqref{equation:PL} and the parameter settings are
from~\cite{a-b},
which are summarized in Table~\ref{tab:plm}.

\begin{table}[!t]
 \caption{\label{tab:parameters} Simulation Parameters}
 \centering
 \begin{tabular}{lll}
  \toprule
  Parameter&Symbol&Value\\
  \midrule
  carrier frequency 1&$f_1$&28\,GHz\\
  carrier frequency 2&$f_2$&60\,GHz\\
  28GHz bandwidth&$W^{mm}_1$&850\,MHz\\
  60GHz bandwidth&$W^{mm}_2$&1500\,MHz\\
  Height of BS&$h_{BS}$&10\,m\\
  Height of MR&$h_{MR}$&2.5\,m\\
  Height of UAV&$h_{UAV}$&100\,m\\
  Number of BS transmit antennas&$N_{B_t}$&3\\
  Number of UAV transmit antennas&$N_{U_t}$&3\\
  transceiver efficiency factor&$\eta$&0.5\\
  background noise&$N_0$&-134dBm/MHz\\
  slot time&$\Delta t$&$18\mu s$\\
  beacon period&$T_s$&$850\mu s$\\
  Maximum antenna gain&$G_0$&$20dBi$\\
  Maximum attenuation of antenna model&$A_m$&$26dB$\\
  Half-power beamwidth&$\theta_{-3dB}$&$15^\circ$\\
  \bottomrule
 \end{tabular}
\end{table}

\begin{table}
\caption{Extracted Parameters of the Path Loss Model for Urban Straight Route Scenarios}
\label{tab:plm}
\centering
\begin{tabular}{|c|c|c|c|c|c|}
        \hline
	\multirow{2}{*}{Straight} & \multicolumn{2}{|c|}{$\alpha$} & \multicolumn{2}{|c|}{$\beta$}& \multirow{2}{*}{$\sigma$} \\
	\cline{2-5}
        & $d<=153.3$ & $d>153.3$ & $d<=153.3$ & $d>153.3$ &\\
        \hline
	Urban & 108.75 & 42.34 & -1.45 & 1.59 & 5.85 \\
	\hline
\end{tabular}
\end{table}

For the performance evaluation study, the following performance metrics are used.

\begin{itemize}
  \item Number of completed flows: the number of scheduled flows with their QoS requirements
  satisfied. If the QoS requirement cannot be met, the
  flow that has
  been scheduled will
  not be
  counted as a completed flow.

  \item System throughput: the total throughput of the entire
  network
  system. This metric is the sum
  of the
  average throughputs
  of all the flows carried
  in the network.

  \item Total amount of time slots: the total number of time slots used to transmit
  all the scheduled flows.
  The actual value of this metric must be smaller than the total
  number of TSs in the frame.
\end{itemize}

For
comparison purpose, we implement the state-of-the-art QoS-aware Concurrent Scheme~\cite{QD} and Maximum QoS-aware
Independent Set (MQIS) based scheduling algorithm~\cite{QoS} as
baseline
schemes.

\begin{itemize}
  \item QoS-aware Concurrent Scheme~\cite{QD}:
  this is a
  concurrent scheduling scheme
  aiming to satisfy the QoS requirements of flows and
  to maximize the number of flows. If
  a flow cannot meet the current scheduling conditions, the scheme cannot relay
  the flow and has to give up scheduling it.

  \item MQIS~\cite{QoS}: this is the scheduling algorithm based on the concept of maximum QoS-aware independent set proposed in~\cite{QoS}. The algorithm divides flows into multiple independent sets. Only after all
  the
  flows in one set are
  scheduled transmitted, the flows in another set can begin to be scheduled.
\end{itemize}

\subsection{Comparison with Existing Schemes} \label{S6-2}

\begin{figure}[t!]
  \begin{center}
  \includegraphics[width=3.4in]{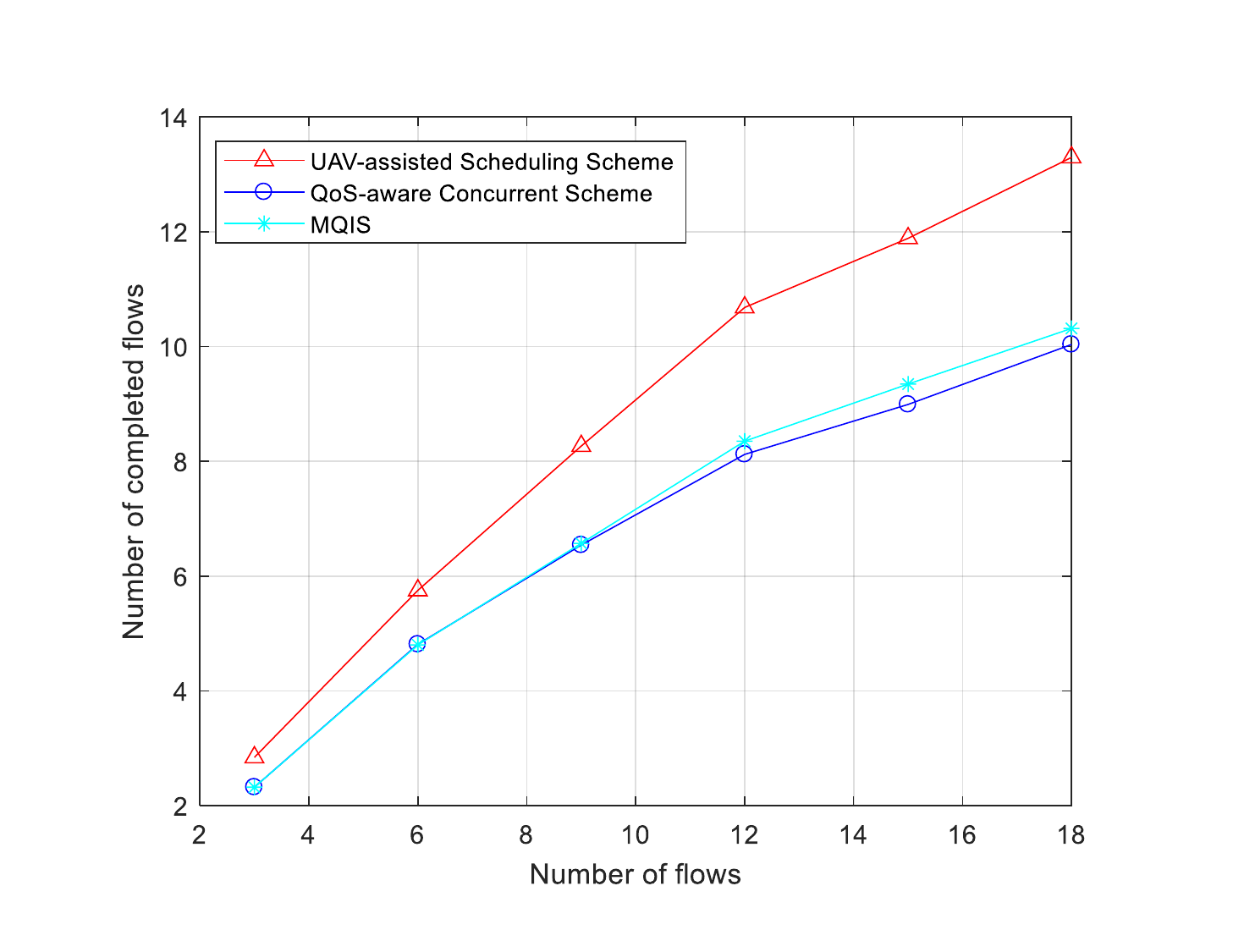}
  \end{center}
  \caption{Number of completed flows versus different numbers of requested flows.} \label{fig:f-f}
\end{figure}

In Fig.~\ref{fig:f-f}, we plot the number of completed flows under different numbers of requested flows. The horizontal distance between the UAV and BS is set to 150$m$, and the number of time slots is set to 8$,$000. With the increased number of requested flows, we can see that
the curves for the three schemes are all rising. The more requested flows in the HSR network, the more flows can complete their scheduling. For the proposed UAV-assisted scheduling scheme, some flows which cannot be directly transmitted from
the
BS
can choose
to be scheduled by UAV relaying. Therefore, the proposed scheme
achieves
a superior performance with
respect to
the number of completed flows compared with
the
other two schemes.
When the number of flows is 18, the proposed scheme
outperforms the QoS-aware concurrent scheme with a gain of
35.0\%
and MQIS with a gain of 32.4\%.

\begin{figure}[!t]
  \begin{center}
  \includegraphics[width=3.4in]{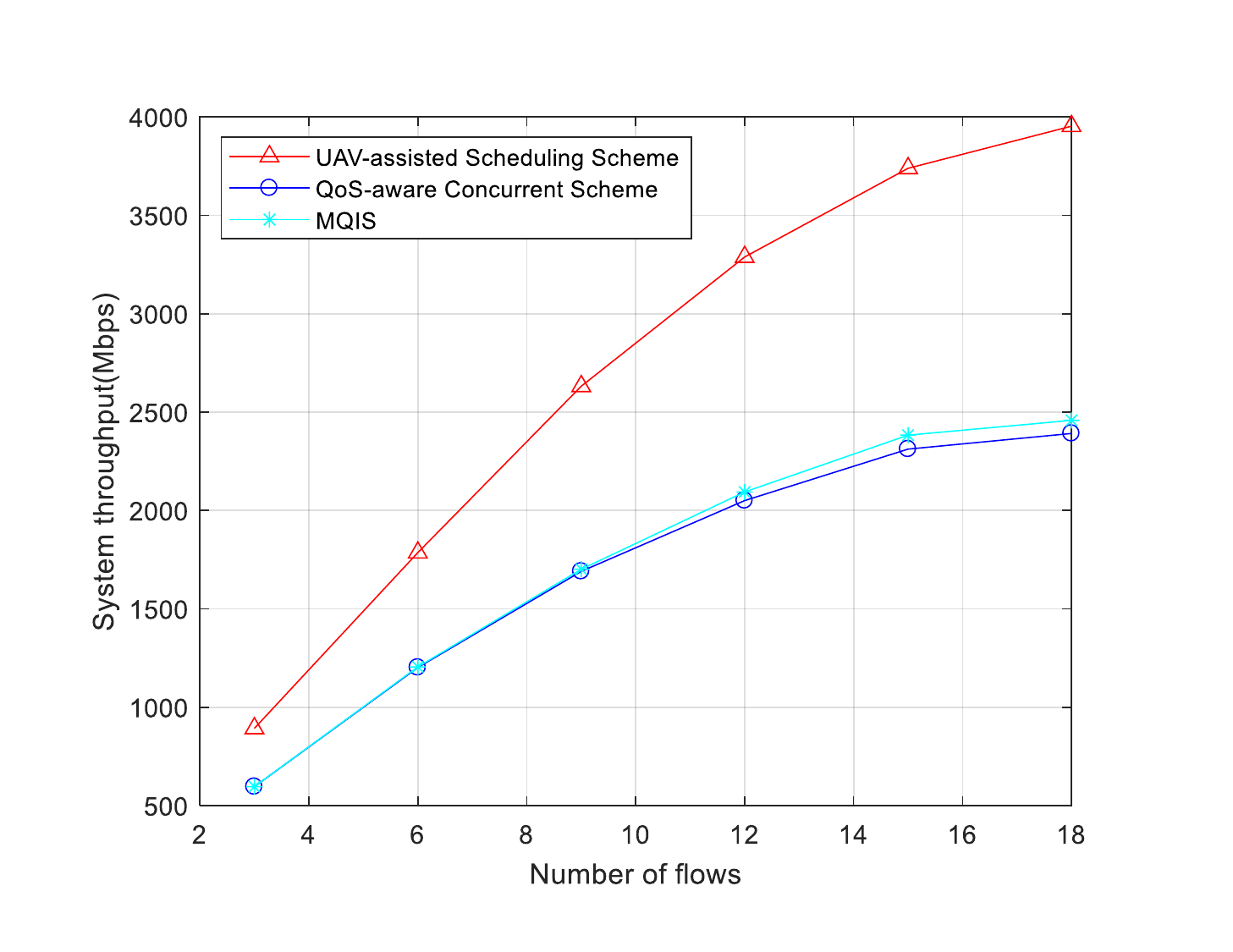}
  \end{center}
  \caption{System throughput versus different numbers of requested flows.} \label{fig:f-t}
\end{figure}

In Fig.~\ref{fig:f-t}, we plot the system throughput under different numbers of requested flows. The horizontal distance between the UAV and BS and the number of time slots are the same as in
Fig.~\ref{fig:f-f}. From the figure, we can see that the system throughput of
the
three schemes are all increasing with the increased number of requested flows. For each flow, our proposed scheme can decide whether to use UAV relaying according to the principle of transmission rate
maximization.
The proposed scheme
achieves the largest system throughput among the three scheme.
It can also relay some flows which cannot be directly
transmitted from the BS. The QoS-aware scheme and
MQIS have similar performance on the system throughput. When the number of requested
flows is 18, the proposed UAV-assisted scheduling scheme
outperforms the QoS-aware concurrent scheme with a gain of
64.6\%
and MQIS with a gain of 61.2\%.

\begin{figure}[!t]
  \begin{center}
  \includegraphics[width=3.4in]{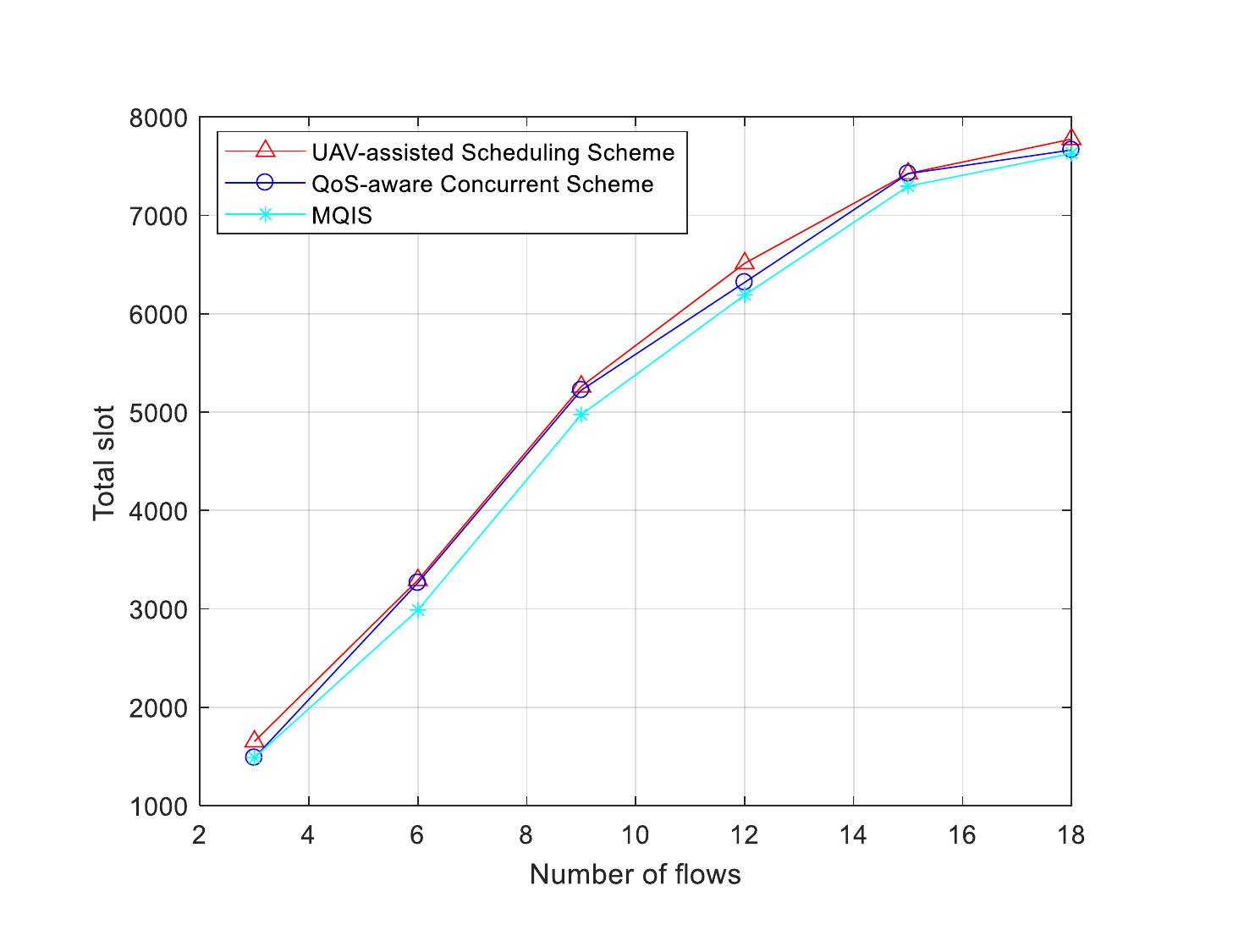}
  \end{center}
  \caption{Number of
  time slots versus different numbers of requested flows.} \label{fig:f-s}
\end{figure}

In Fig.~\ref{fig:f-s}, we plot the system throughput under different numbers of requested flows. The horizontal distance between the UAV and BS and the number of time slots
remain the same as previous simulations.
From this figure, we can see that
the total number of
slots of
the
three schemes are all increasing as
more flows are requested.
Although
the proposed scheme can schedule the most flows among the three schemes, as shown in Fig.~\ref{fig:f-f},
it can be seen from Fig.~\ref{fig:f-s} that the total numbers of time slots needed by the three schemes are very similar. This proves that the proposed UAV-assisted scheme has a superior performance on
transmission efficiency compared with the QoS-aware concurrent scheme and
MQIS.

\begin{figure}[!t]
  \begin{center}
  \includegraphics[width=3.4in]{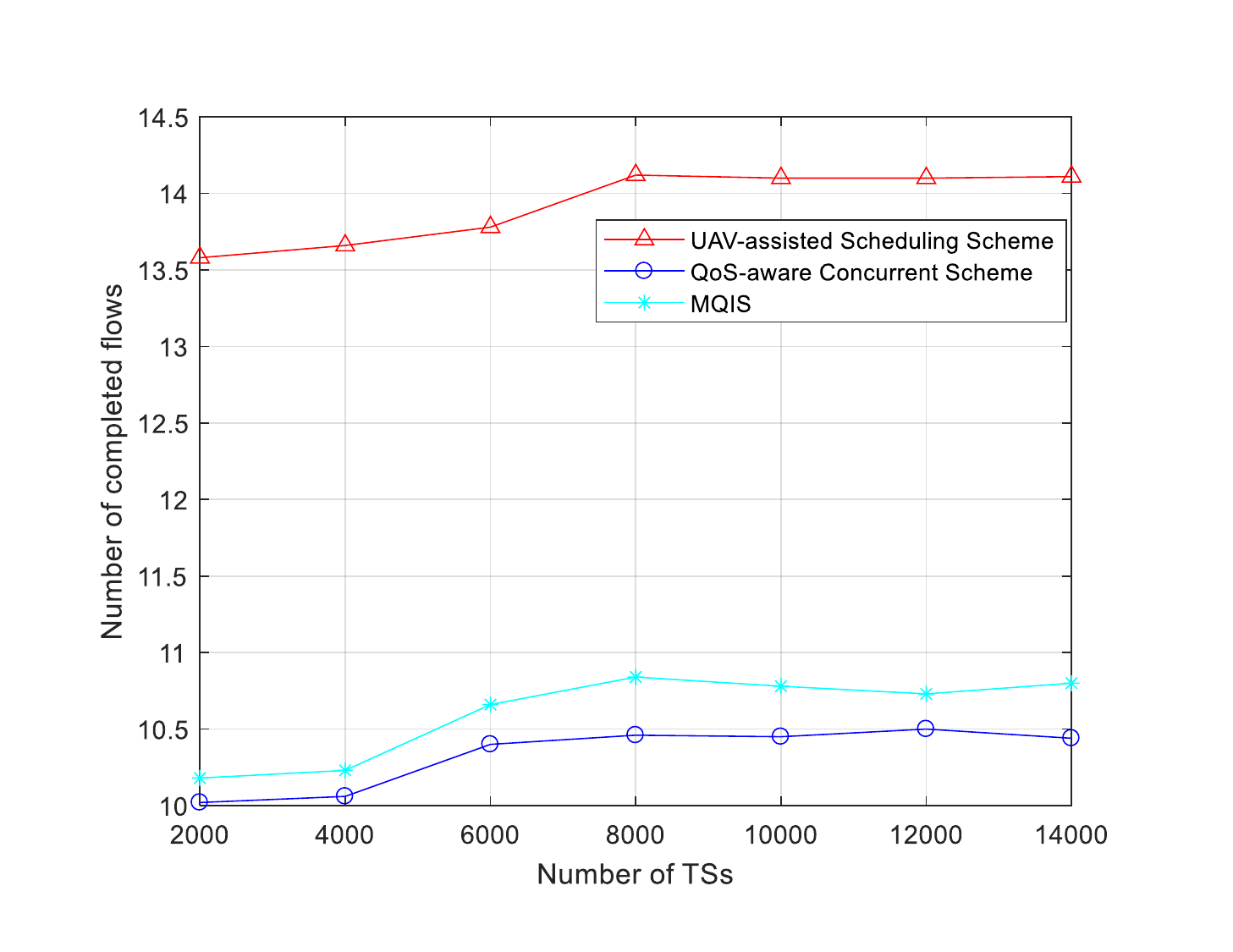}
  \end{center}
  \caption{Number of completed flows versus different numbers of time slots in each frame.} \label{fig:s-f}
\end{figure}

In Fig.~\ref{fig:s-f}, we plot the number of completed flows under different numbers of time slots
in
each frame. The horizontal distance between the UAV and BS is set to 150 m, and the number of requested flows is set to 18. When the number of time slots
begin
to increases, the
curves of the three schemes are all rising. When the number of time slots increases to a certain value, the numbers of completed flows of three schemes are all becoming steady. Therefore, the number of time slots does not need to be increased to a very large value;
an appropriate value of time slot number in each scheme can
ensure a satisfactory performance.

\begin{figure}[!t]
  \begin{center}
  \includegraphics[width=3.4in]{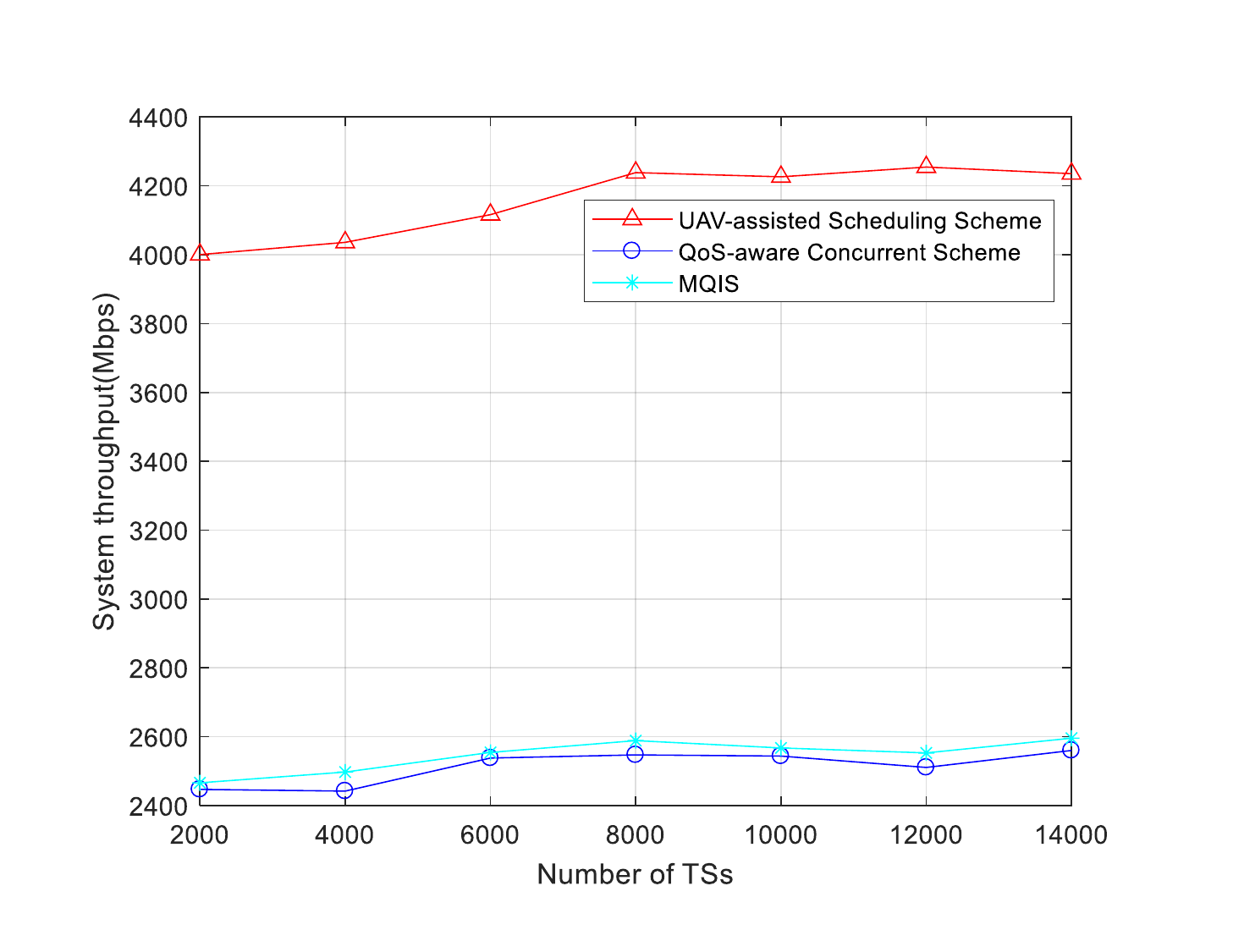}
  \end{center}
  \caption{System throughput versus different numbers of time slots.} \label{fig:s-t}
\end{figure}

In Fig.~\ref{fig:s-t}, we plot the system throughput under different numbers of time slots. The horizontal distance between UAV and BS and the number of requested flows are the same as that in Fig.~\ref{fig:s-f}. From the figure, we can see
similar
trends for the three schemes as in
Fig.~\ref{fig:s-f}. With the increased number of
time slots, we can see that the curves for the three
schemes are all rising. The more time slots, the more system throughput can achieve during scheduling.
The three curves start to stabilize when the number of slots is increased over 8$,$000. Hence, the number of TSs is set to 8$,$000 in the other simulations.

\begin{figure}[!t]
  \begin{center}
  \includegraphics[width=3.4in]{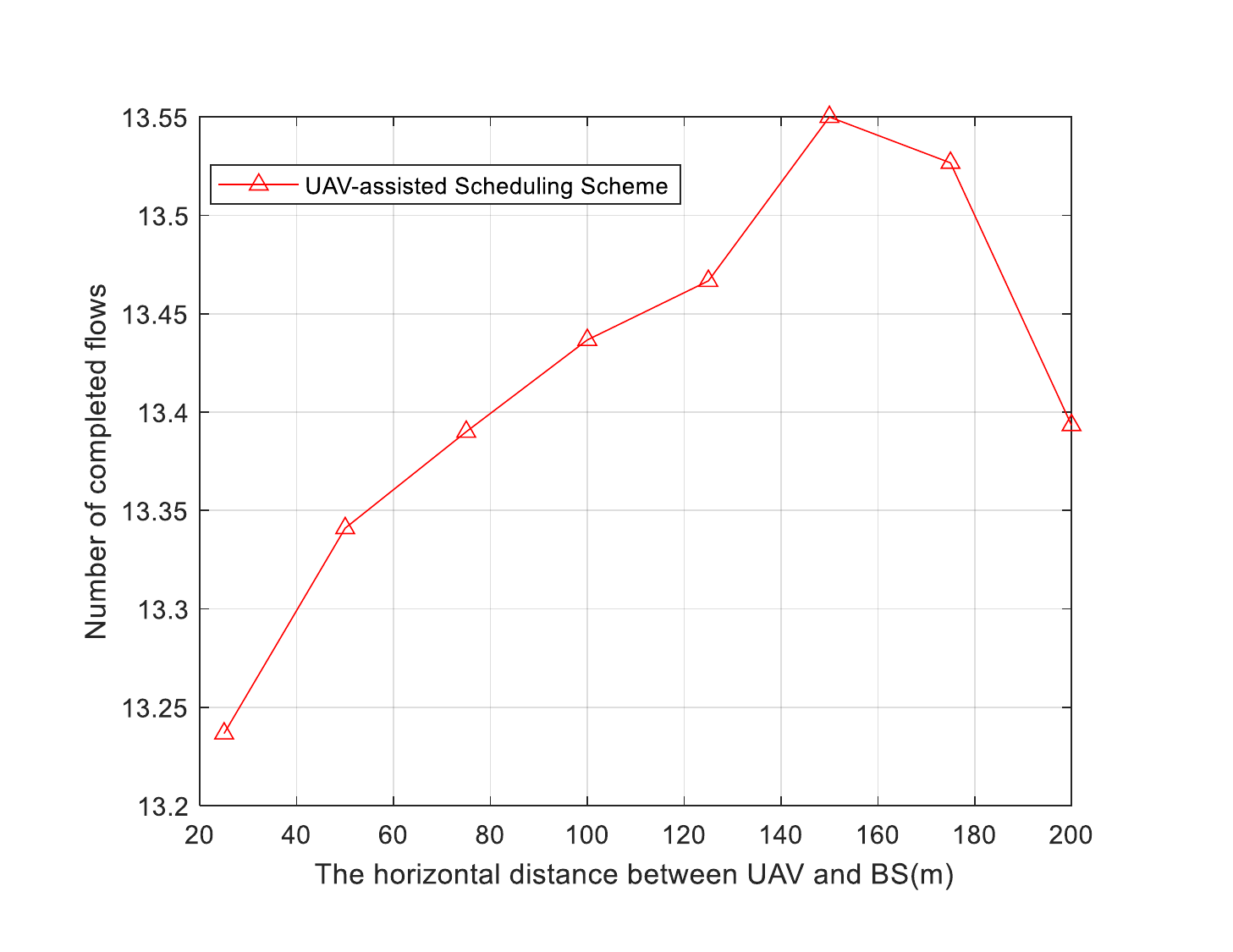}
  \end{center}
  \caption{Number of completed flows versus different horizontal distances between the UAV and BS.} \label{fig:u-f}
\end{figure}

In Fig.~\ref{fig:u-f}, we plot the number of completed flows under different horizontal distances between UAV and BS,
to demonstrate the impact of the UAV location on performance.
The number of requested flows is set to 18, and the number of time slots is set to 8$,$000. With the increased horizontal UAV-BS distance,
the number of completed flows increases first and then declines.
Thus a suitable location of the UAV has a significant impact on the system performance. When
UAV-BS distance is 150 m, the number of completed flows reaches the maximum value. Thus, the horizontal distance between the UAV and BS is set to 150 m in all the other simulations.

\section{Conclusions}\label{S7} 

In this paper, we considered the problem of scheduling flows with diverse QoS requirements in the scenario of
mmWave HSR communications. To maximize the number of flows
while satisfying their QoS requirements,
we proposed the UAV-assisted scheduling scheme to deploy a UAV relay to assist the scheduling of flows.
Our simulation study showed that the
proposed UAV-assisted scheme outperformed
two baseline schemes on the
number of completed flows, achievable system throughput, and transmission efficiency.

\vfill

\end{document}